\documentclass[runningheads]{llncs}
\usepackage[nounderscore]{syntax}
\usepackage{graphicx}
\usepackage{booktabs}
\usepackage{listings}
\usepackage{csquotes}
\usepackage{subcaption}
\usepackage{svg}
\usepackage{stmaryrd}
\usepackage{amsmath}
\usepackage{amsfonts}
\newcommand{\sema}[1]{{\llbracket {#1} \rrbracket}}

\usepackage{comment}
\usepackage{fancyvrb}
\usepackage{color}
\usepackage[breaklinks=true]{hyperref}

\newcommand{\LL}{\mathcal{L}}
\newcommand{\PP}{\mathcal{P}}
\newcommand{\II}{\mathcal{I}}
\newcommand{\CC}{\mathcal{C}}

\newcommand{\etal}{\textit{et al.}}

\definecolor{codegreen}{rgb}{0,0.6,0}
\definecolor{codegray}{rgb}{0.5,0.5,0.5}
\definecolor{codepurple}{rgb}{0.58,0,0.82}
\definecolor{backcolour}{rgb}{0.94,0.94,0.95}

\usepackage{xspace}
\newcommand{\sys}{TorchProbe\xspace}
\usepackage{xcolor}

\begin{document}
\input{code/fmt} % needed for defining code format macros
\title{\sys : Fuzzing Dynamic Deep Learning Compilers}

\author{Qidong Su\inst{1,2}\and
Chuqin Geng\inst{1,3,4}\and
Gennady Pekhimenko\inst{1,2}\and
Xujie Si\thanks{Corresponding author}\inst{1,2,4}
}

\authorrunning{Q. Su et al.}

\institute{University of Toronto\and
Vector Institute\and
McGill University\and
Mila - Quebec AI Institute \\
\email{\{qdsu, pekhimenko, six\}@cs.toronto.edu\\chuqin.geng@mail.mcgill.ca}
}

\maketitle              % typeset the header of the contribution
\begin{abstract}

Static and dynamic computational graphs represent two distinct approaches to constructing deep learning frameworks. The former prioritizes compiler-based optimizations, while the latter focuses on programmability and user-friendliness. The recent release of PyTorch 2.0, which supports compiling arbitrary deep learning programs in Python, signifies a new direction in the evolution of deep learning infrastructure to incorporate compiler techniques in a more dynamic manner and support more dynamic language features like dynamic control flows and closures. Given PyTorch's seamless integration with Python, its compiler aims to support arbitrary deep learning code written in Python. However, the inherent dynamism of Python poses challenges to the completeness and robustness of the compiler. While recent research has introduced fuzzing to test deep learning compilers, there is still a lack of comprehensive analysis on how to test dynamic features. To address this issue, we propose several code transformations to generate test cases involving dynamic features. These transformations preserve the program's semantics, ensuring that any discrepancy between the transformed and original programs indicates the presence of a bug. Through our approach, we have successfully identified twenty previously unknown bugs in the PyTorch compiler and its underlying tensor compiler Triton.

% \keywords{First keyword  \and Second keyword \and Another keyword.}
\keywords{Debugging  \and Software engineering \and Deep learning}

\end{abstract}
\section{Introduction}
% DL is popular. 
% DL needs performance optimization.
% DL compiler is one of the most important techniques.
% Deep Learning (DL) has recently achieved significant success in many important fields like vision, natural language processing, robotics, autonomous driving, etc. Numerous algorithms, model architectures, and applications are emerging nowadays, requiring highly flexible infrastructure.

% People use python for DL
% As a programming language with highly dynamic features, Python has become the de facto standard in the DL community for its expressiveness, flexibility, and ecosystem. The most popular DL frameworks like TensorFlow~\cite{tensorflow} and PyTorch~\cite{pytorch} all provide their programming interface as a domain-specific language (DSL) embedded in Python.

Deep Learning (DL) has recently achieved significant success in various critical fields such as vision~\cite{resnet,vit}, natural language processing~\cite{bert,instructgpt,gpt3}, and autonomous driving~\cite{cordts2016cityscapes}. This progress has led to the emergence of numerous algorithms, model architectures, and applications, necessitating highly flexible infrastructure. Python, with its highly dynamic features, has become the de facto standard programming language in the DL community. It is favored for its expressiveness, flexibility, and rich ecosystem. Popular DL frameworks like TensorFlow~\cite{tensorflow} and PyTorch~\cite{pytorch} provide their programming interfaces as domain-specific languages (DSLs) embedded in Python.

However, executing high-level DL code efficiently in Python remains a challenging problem. Training and inference of deep learning models are known to be time and resource-consuming, particularly given the increasing number of parameters. Hardware vendors offer various accelerators such as GPUs~\cite{amd,gpu}, TPUs~\cite{tpu}, and NPUs~\cite{cambricon,trainium,graphcore,cerebras,sambanova,inferentia} to address this issue. These accelerators typically have distinct specifications, architectures, and programming models. Therefore, effectively mapping DL tasks onto the underlying hardware is crucial for maximizing the utilization of accelerators.

To bridge the gap between abstract deep neural network descriptions and the low-level hardware instructions of accelerators, the concept of "deep learning compilers"\cite{mlc-survey} has been introduced. The typical workflow of a deep learning compiler involves three steps: 1) writing high-level DL code in Python, 2) converting the DL model into an intermediate representation (IR) provided by the compiler, and 3) calling the DL compiler to generate optimized code. However, while Python itself is a highly flexible language, the IRs in DL compilers are usually more restricted as they rely on compile-time information to enable more optimization opportunities. For instance, static computational graphs allow for more advanced graph-level optimizations e.g., operator fusion~\cite{dnnfusion,tf-graph}. Another example is fixing the shape of tensors involved in computations, which facilitates finding optimal configurations for the generated code, such as tiling size~\cite{autotvm}.

The misalignment in expressiveness between Python and deep learning compilers poses challenges and requires more manual intervention and engineering effort, hindering the widespread application of compiler techniques. TorchScript~\cite{torchscript}, the first-generation compiler of PyTorch, provides an intermediate representation (IR) that is incompatible with Python. As a result, users are faced with the options of either rewriting their code in a constrained subset of Python (scripting) or tracing the code's execution trajectory, which only captures partial information of the original code (tracing). On the other hand, Torch.fx~\cite{torch.fx}, the second-generation compiler, offers an IR compatible with Python but requires the code to be purely functional.

In response to this issue, a recent trend in deep learning compiler development is to support more dynamic features, enabling the seamless application of compiler techniques within the original Python language. The newly released PyTorch 2.0~\cite{pt2} includes a compiler component that facilitates the automatic optimization of \emph{any} Python code through a simple API called \texttt{torch.compile}. This compiler component modifies the process of launching a Python function by analyzing the bytecode generated by the Python interpreter. It identifies code snippets that can be optimized, performs compiler optimizations on them, and caches the optimized code for future reuse. Given that deep learning tasks often involve high levels of repetition, this strategy can yield significant performance improvements in many scenarios.

However, the implementation of the new PyTorch compiler is complex and intertwined with the original Python interpreter, DL frameworks and libraries, and heterogeneous hardware. This complexity introduces the risk of potential bugs in the implementation, many of which may remain hidden and only manifest in specific corner cases. Manually creating test cases as a solution requires significant engineering effort and can only cover a limited range of possible inputs. Therefore, an automated testing framework that generates test cases for dynamic deep learning compilers would expedite the process of bug detection and enhance the robustness of the infrastructure. While previous works have utilized \emph{fuzzing} techniques on DL infrastructures to identify bugs, they focus on static computational graphs~\cite{nnsmith,audee,lemon,graphfuzzer,muffin}, and the exploration of how to effectively fuzz dynamic deep learning compilers is still an area that lacks comprehensive analysis, to the best of our knowledge.

To address these challenges, we present a novel fuzzing framework called \sys. This framework is designed to generate test cases that cover dynamic features such as control flows, in-place tensor mutation, list comprehension, and nested functions. For each generated test case, we perform three checks: 1) ensuring that the compiler can generate optimized code without encountering errors, 2) verifying that the optimized code can be executed successfully, and 3) validating that the output produced by the optimized code remains consistent with the original program.

To ensure the meaningfulness of the tests, it is crucial that they are both logically and numerically valid, capable of producing valid output even before being optimized by the compiler. For instance, the test cases should be free of numerical errors (e.g., division by zero) and undefined behaviors like \texttt{INFINITY} and \texttt{NaN} in floating-point numbers. Furthermore, the introduction of control flows can potentially lead to programs that never terminate, such as infinite loops.

To ensure the validity of the test cases, we employ two types of program mutation to generate new test cases based on a ``seed" program: \emph{Equivalent Mutation} and \emph{Equivalence Modulo Inputs} (EMI)~\cite{sun2016finding}. We ensure that the test cases generated through these mutations always produce the same output as the seed program if the compiler processes them correctly. In our approach, we convert the computational graphs generated by an existing fuzzer NNSmith~\cite{nnsmith} into straight-line code, which serves as the seed program. Since the methods for guaranteeing the validity of computational graphs have been extensively studied in prior works~\cite{graphfuzzer,nnsmith}, we can ensure the validity of the mutated programs.

\begin{figure}[h]
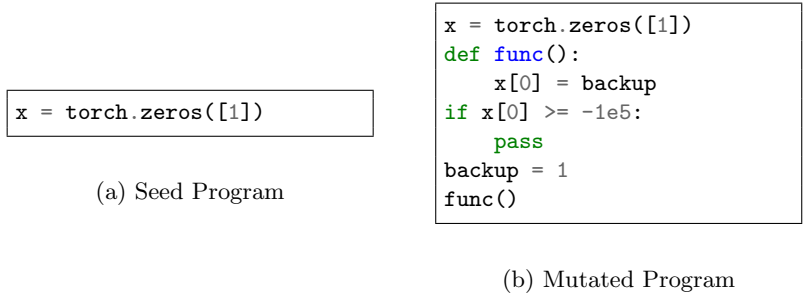

\centering
\hspace{20pt}
\begin{subfigure}[c]{0.4\textwidth}
\input{code/intro-exampl1.py}
\caption{Seed Program}
\end{subfigure}\hfill
\begin{subfigure}[c]{0.4\textwidth}
\input{code/intro-exampl2.py}
\caption{Mutated Program}
\end{subfigure}
\hspace{20pt}
\caption{An example of constructing more sophisticated test cases from a seed program. This test case triggers the PyTorch compiler to crash because it does not handle hoisting well.}
\label{fig:example}
\end{figure}

Moreover, these mutations are composable. Each mutation introduces more dynamic language features, and their composition enlarges the detected program space, which potentially exposes more bugs. A simple illustrative example is as follows in Figure~\ref{fig:example}, where we derive a test case from one line of tensor declaration via several steps of program mutations. PyTorch compiler crashed on this program because it failed to handle hoisting, closures, and graph breaks correctly.  One core developer of PyTorch refers to this bug as \emph{`a lot of fun for a PL nerd'}.

The program mutations presented in this paper represent merely \emph{one potential methodology} for generating valid test cases. It's important to note the validity of test cases does not necessarily rely on the equivalence. While equivalence-based program mutations can generate valid test cases, there exist huge numbers of valid test cases that are not covered by it. This work did not investigate how to synthesize valid test cases beyond equivalent program mutations, which is a vast space to explore for future works.

We summarize the main contributions of this work as follows:
\begin{enumerate}
    \item We observe the problem of how to automatically detect bugs in the emerging dynamic deep learning compilers, which aim to support more dynamic language features than previous ones.
    \item We design novel program mutations to automatically generate test cases specifically for dynamic language features such as control flows, data mutation, and closure.
    \item We find twenty previously undiscovered bugs in the latest compiler of PyTorch and its underlying backend compiler Triton. All of them are confirmed by the community, twelve of them have already been fixed, and five test cases have been integrated as unit tests. 
    % \item \qd{TODO: coverage}
\end{enumerate}

\section{Background}

In this section, we introduce related background including the recent development of deep learning compilers and automatic testing techniques for deep learning systems.

\subsection{Deep Learning Program and Systems}
Deep learning applications differ from generic programs in several aspects, with the most significant one being their high computation intensity. DL code often involves heavy computations such as matrix multiplications and convolutions, which require a large number of floating-point operations. Running unoptimized DL tasks on regular hardware leads to unacceptable slowness.

To tackle this challenge, a domain-specific hardware-software stack has been established, encompassing accelerators, libraries, runtimes, frameworks, and languages. Hardware accelerators like GPUs, TPUs, and NPUs offer beneficial features for DL applications, such as highly parallel SIMT programming models and dedicated circuits for matrix multiplications~\cite{tensorcore}. However, new hardware also poses challenges for upper-level software. Developing high-performance code that is optimized for specialized hardware demands expertise in computer architecture and significant engineering effort. To address this, numerous solutions have been proposed.

% Deep learning applications differ from generic programs in many aspects, among which the most significant one is the high computation intensity. DL code often contains heavy computations such as matrix multiplications and convolutions, which have huge numbers of floating point operations. Running unoptimized DL tasks on ordinary hardware is unacceptably slow.

% To address this, a domain-specific hardware-software stack has been established, from accelerators, libraries, runtimes, frameworks, and languages. Hardware accelerators such as GPUs, TPUs, and NPUs have many features beneficial for DL applications, like highly parallel SIMT programming models or dedicated circuits for matrix multiplications~\cite{tensorcore}. However, new hardware also brings challenges for upper-level software. Writing high-performance code optimized for specialized hardware requires expertise in computer architecture and heavy engineering effort. To address this, many solutions have been proposed.

\paragraph{High-performance Libraries} One solution is to use high-performance math libraries provided by hardware vendors. Hardware vendors often provide highly specialized and optimized libraries for their hardware products, such as cuBLAS~\cite{cublas} and cuDNN~\cite{cudnn} for NVIDIA GPUs, and MKL-DNN~\cite{mkl} for Intel CPUs. These libraries contain many commonly used operators in the form of parallel functions called \emph{kernels}. Most of them are written in C/C++ or assembly languages to provide high performance.

\paragraph{Deep Learning Frameworks} While high-performance libraries provide superior performance, their programming interface is not sufficiently friendly to DL researchers. Therefore, deep learning frameworks such as TensorFlow and PyTorch are invented for better user interface and programmability. A common abstraction of DL frameworks is \emph{computational graphs}, which we elaborate on in Section~\ref{sec: comp-graph}.

% \paragraph{Programming Languages for DL} While there are many domain-specific programming languages designed for DL tasks, Python is the most popular one. Almost all DL frameworks heavily rely on Python or provide Python interface.

\subsection{Computational Graphs}\label{sec: comp-graph}
% Computational graphs are a commonly used abstraction for writing DL programs, automatic differentiation, and performance optimization. DL programs can be viewed as graphs of data flow, where each node represents an operator and the edges represent data dependencies.

% Computational graphs can be categorized into two types, namely static and dynamic graphs, according to when and how they are built.

Computational graphs are a widely used abstraction in deep learning programs, providing benefits in automatic differentiation and performance optimization. These graphs represent the data flow of DL programs, with nodes representing operators and edges representing data dependencies.

There are two types of computational graphs: static and dynamic. They are categorized based on when and how they are constructed.

\paragraph{Static Graph} Many frameworks such as Caffe~\cite{caffe}, Theano~\cite{theano}, and early versions of Tensorflow~\cite{tensorflow} adopt static computation graphs. When writing DL programs, users explicitly describe the computational graph using the primitives provided by the framework. The computational graph is not executed until it is fully constructed (or \emph{define-then-run}). The advantage of static graphs is that the framework can obtain complete information about the computational graph before the program is executed, which enables optimizations based on graph analysis~\cite{tf-graph}, such as constant folding, common sub-expression elimination, and operator fusion. 

\paragraph{Dynamic Graph} While static graphs provide opportunities for various optimizations, the define-then-run programming interface is not aligned with the imperative programming paradigm and other well-known matrix libraries such as \texttt{numpy}~\cite{numpy}, which imposes additional demands on users. PyTorch~\cite{pytorch} and TensorFlow Eager~\cite{tf-eager} adopt another strategy called dynamic graphs. In PyTorch, the code of forward propagation is similar to imperative numpy code. When an operator is called, its computation will be triggered immediately, and it will also be book-kept by a \emph{tape}, which constructs the computational graph on the fly (or \emph{define-by-run}). The computational graph is later used in the backward propagation. While dynamic graphs provide a better programming interface, it makes performance optimizations more challenging~\cite{torch.fx}.

\begin{figure}
\centering

\begin{subfigure}[t]{\linewidth}
\centering
\includegraphics[width=0.6\textwidth]{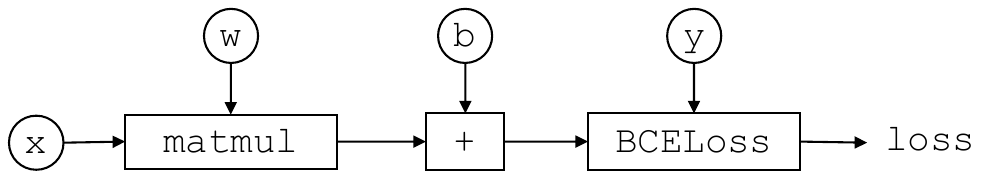}
\end{subfigure}

\begin{subfigure}[t]{0.8\linewidth}
\input{code/bg-example.py}
\end{subfigure}

\caption{An example of computational graph and its corresponding PyTorch code.}
\label{fig:enter-label}
\end{figure}

\subsection{Deep Learning Compilers}
% \qd{the motivation of using compilers}

% \qd{should double check torch.fx paper}
Compilers play an important role in many layers of the DL software stack. The CUDA compiler \texttt{nvcc} compiles CUDA code into lower-level binary code. Tensor compilers such as Halide~\cite{halide}, TVM~\cite{tvm}, and Hidet~\cite{hidet} translate high-level mathematical tensor expressions into efficient kernels for different target hardware. This process is called \emph{scheduling}, which is finished either manually or automatically. Triton is an intermediate language between CUDA and other loop-based tensor compilers, providing a programming model at the tile level.

Generally speaking, most DL compilers work on two distinct levels, namely the graph level and operator (or kernel) level. They take computational graphs as inputs, divide them into smaller subgraphs, and generate efficient kernels for each subgraph. 

However, there is a misalignment between the language which DL researchers use and the frontend language used by the DL compilers. While a lot of DL researchers write programs in Python, most DL compilers take static graphs as their input. So it is necessary to convert the DL models to formats that can be accepted by DL compilers. Some examples of these formats are ONNX~\cite{onnx} and TorchScript~\cite{torchscript}. The conversion is far from seamless due to different operator sets and implementation details.

The PyTorch community provides several solutions to ease the process of translating PyTorch code into compilable formats. TorchScript~\cite{torchscript} is a language designed for DL model deployment. There are two methods to convert PyTorch code into TorchScript namely scripting and tracing. Scripting requires users to write in a subset of Python, which will be later parsed and translated into TorchScipt directly. Tracing is to actually run the code and record which operators are launched, and the recorded trace is later translated into TorchScript. Both scripting and tracing have weaknesses. Scripting does not support all language features of Python so users often need to rewrite the code, while tracing loses many key information such as control flows.

\texttt{Torch.fx}~\cite{torch.fx} is purely functional language designed for compiling PyTorch code, which is equivalent to computational graphs. Users can translate PyTorch code into torch.fx via symbolic tracing. However, many language features are still not supported. Figure~\ref{fig:fxfail} shows two examples that torch.fx cannot handle. 
\begin{figure}\centering
\hfill
\begin{subfigure}[c]{0.4\linewidth}
\input{code/bg-fx-fail.py}
\caption{Tensor mutation}
\end{subfigure}\hfill
\begin{subfigure}[c]{0.4\linewidth}
\input{code/bg-fx-fail2.py}
\caption{Dynamic control flows}
\end{subfigure}
\hfill
\caption{Two examples of PyTorch code that torch.fx fails to trace, as 
it is a purely functional language design.}
\label{fig:fxfail}
\end{figure}

\subsection{Dynamic Deep Learning Compilers}
% \qd{More about dynamism in DL. Why do we need dynamism in DL?}

\begin{figure}
    \centering
    \includegraphics[width=\linewidth]{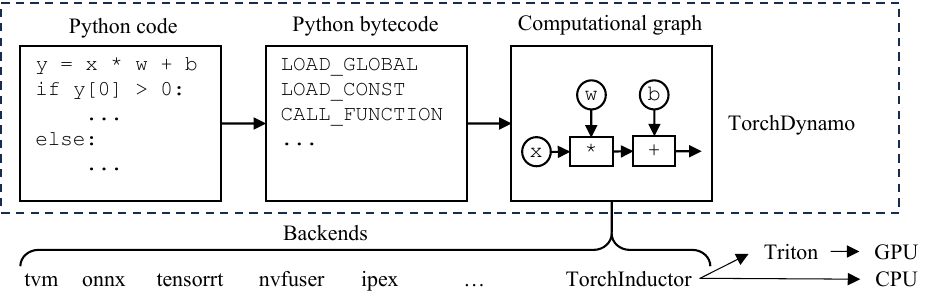}
    \caption{Overview of the compiler of PyTorch 2.0 -- TorchDynamo. TorchDynamo analyzes bytecode and captures partial computational graphs. It translates these graphs into torch.fx graphs, which can be compiled by backend compilers like TensorRT in a Just-in-Time manner. }
    \label{fig:pt2}
\end{figure}

The recently released PyTorch 2.0~\cite{pt2} adopts a new strategy of compilation, which is called \emph{TorchDynamo}. 
The workflow of the PyTorch compiler is shown in Figure~\ref{fig:pt2}. 
While computational graphs cannot represent all PyTorch programs, TorchDynamo analyzes the bytecode generated by the Python interpreter and capture \emph{partial} computational graphs and translate them into torch.fx graphs.
Therefore, a Python function might be broken down into multiple partial graphs, and the parts which cannot be compiled will be executed as normal Python code by the interpreter. 
In this manner, TorchDynamo has the ability to support \emph{arbitrary} Python code.

Captured computational graphs will be compiled by one of the \emph{backend} compilers in a Just-in-time way and cached for future reuse. Since DL tasks are repetitive, the compiled code has a high chance to be reused. Every time one function is called, TorchDynamo will check whether it has already been compiled and the conditions of reuse are satisfied (called \emph{guards}, e.g. the shape of the input tensors is not changed), and will execute the compiled version if all requirements are met, which leads to substantial speedup.

TorchDynamo now supports many backend compilers including TorchInductor, TVM~\cite{tvm}, TensorRT~\cite{tensorrt}, and ONNX~\cite{onnx}. The official backend compiler is TorchInductor, which can generate high-performance CPU kernels, or relies on the Triton~\cite{triton} intermediate language to generate GPU kernels.

\subsection{Fuzzing Deep Learning Systems}
As deep learning is applied to many critical scenarios, how to guarantee the robustness of deep learning systems has become an important topic. Fuzzing as an automatic testing technique has been introduced to detect bugs in DL systems. NNSmith~\cite{nnsmith} generates random valid computational graphs and input data to check if the tested DL system can produce expected outputs. A more detailed discussion about related work is included in Section~\ref{sec: related}.

\section{Approach}\label{sec: appr}

This section presents our proposed deep learning fuzzer for testing dynamic deep learning compilers. 

\subsection{Overview}

\begin{figure}[t]
\centering
\includegraphics[width=\textwidth]{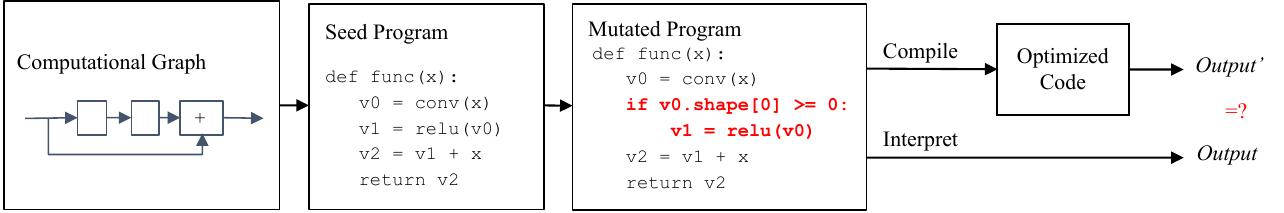}
\caption{The TorchProbe workflow involves translating input computational graphs into Python code, which serves as seed programs. These seed programs undergo numerical-equivalence mutations to create corresponding mutated programs. Finally, we verify if the compiler can produce the same results from the original Python interpreter and eager-mode PyTorch.}
\label{fig:workflow}
\end{figure}

Figure \ref{fig:workflow} illustrates the overall workflow of our fuzzer.
Let $\mathcal{L}$ be the language space of all possible programs, which specifically refers to code written in Python and PyTorch in our case.
Let $\II$ be the input space which includes all possible input data, e.g. tuples of tensors.

A correctly implemented compiler should guarantee that the optimized code should generate the same output as the original code. More formally, let $\sema{\PP}(x)$ be the output of the program $\PP \in \LL$ given the input $x\in\II$, executed on the original Python interpreter and eager-mode PyTorch, 
and $\sema{\PP}_\CC(x)$ be the output of the compiled version of $\PP$ produced by the compiler $\CC$, which is executed on the modified runtime system for compiled code.
The correctness of a compiler could be represented as
\begin{equation}\label{eq:oracle}
\forall \PP\in\LL, x\in\II, \text{ $\sema{P}(x)$ is valid} \Rightarrow \sema{\PP}(x) = \sema{\PP}_\CC(x) 
\end{equation}
where `$\sema{\PP}(x)$ is valid' means the program $\PP$ is executed correctly and produces meaningful results without any syntax or semantic errors, runtime errors, numerical errors, or undefined behaviors including \texttt{INFINITY} and \texttt{NaN} in tensors. The compiler can pass the test case if 1) It can finish the compilation without errors, 2) the compiled code is runnable, 3) the compiled code can produce consistent outputs as the original program.

A fuzzer can sample valid test cases $(\PP, x)$ containing a program $\mathcal{P} \in \mathcal{L}$ and input data $x \in \II$, and check whether the compiled code can produce consistent results as the original code. In order to ensure the validity of samples, we start from a valid `seed' test case and perform a series of numerical-equivalence \emph{mutations}, which preserves its validity in each step. We introduce two categories of mutations, namely \emph{equivalent mutations} and \emph{equivalence module inputs}. Therefore, as long as the seed test case is valid, the validity of derived test cases can also be guaranteed. We also design these mutations to be composable so that the composition of mutations form more sophisticated test cases, which might expose more bugs.

In order to obtain a valid test case, we use NNSmith \cite{nnsmith}, a fuzzer for DL systems based on computational graphs. It can automatically generate valid test cases, including the computational graph and input data. We then translate the computational graphs into straight-line Python code in static single-assignment form (SSA), which we use as the seed program. A seed program consists of a list of statements in the form of 
$$
out = op(in_1, in_2, \dots, in_n)
$$
where $op$ is a DL operator, such as matrix multiplication or convolution, $in_i$-s are input tensors, and $out$ is the output tensor. Each variable will appear as the output in at most one statement, that is, be defined once and remain constant.

\subsection{Equivalent Mutations}
Given a target deep learning program, the most direct approach to diversify the program is through equivalent mutations. 
Equivalent mutations ensure that both the seed program $\mathcal{P}$ and the mutated program $\mathcal{P}^\prime$ yield the same output results for any random input $x$ within the valid domain $\mathcal{I}$:
\begin{equation}
\forall x \in \mathcal{I}, \quad\text{$\sema{P}(x)$ is valid} \Rightarrow \sema{\PP}(x) = \sema{\PP'}(x)
\end{equation}

The validity of the mutated test case $(\PP', x)$ comes from the fact that $\sema{\PP}(x)$ is valid. Compiler bugs can be detected if the compiler crashes or the mutated program $\mathcal{P}^\prime$ produces results that differ from the original program and the reference (oracle) program. In other words, the compiler passes the test if $\sema{\PP'}(x) = \sema{\PP'}_\CC(x)$.

\begin{figure}
    \centering
    \begin{subfigure}{0.68\linewidth}
    \includegraphics[width=\textwidth]{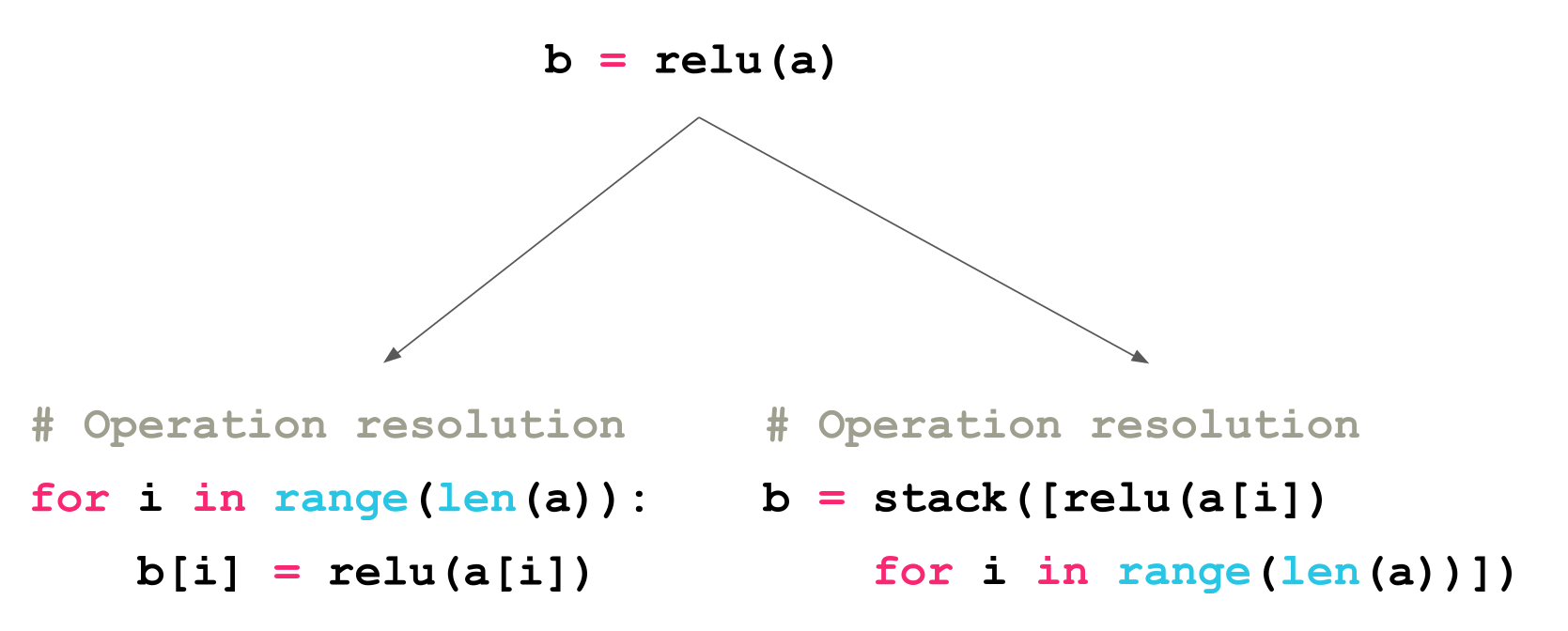}
    \caption{Operator resolution.}
    \label{fig:or}
    \end{subfigure}\\
    \begin{subfigure}{0.3\linewidth}
    % \centering\captionsetup{width=0.96\linewidth}%
    \includegraphics[width=\textwidth]{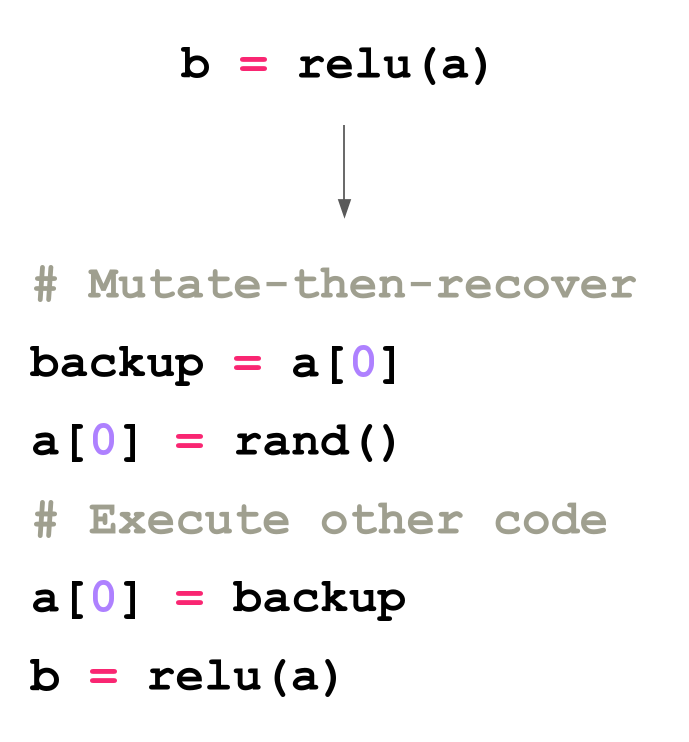}
    \caption{Mutate-then-recover.}
    \label{fig:mtr}
    \end{subfigure}\hspace{60pt}
    \begin{subfigure}{0.27\linewidth}
    % \centering\captionsetup{width=0.96\linewidth}%
    \includegraphics[width=\textwidth]{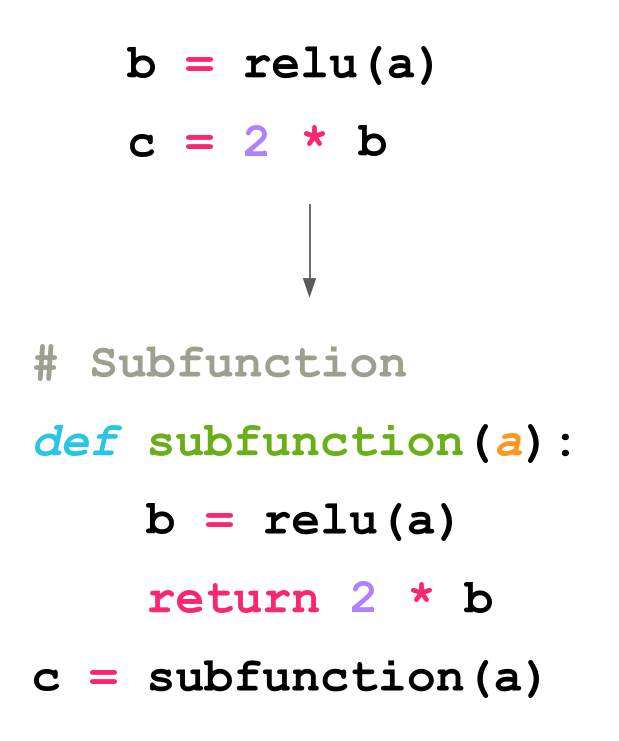}
    \caption{Functionalize}
    \label{fig:subfunc}
    \end{subfigure}
    \caption{The three types of equivalent mutations.}
    \label{fig:eq-mutations}
\end{figure}

Our equivalent mutation method iteratively introduces more dynamic components to the target program. Specifically, we consider three types of mutations (also illustrated in Figure \ref{fig:eq-mutations}):
\begin{enumerate}
\item \textbf{Operator resolution}. Deep learning operators are primarily vectorized operators. These operators can be fully or partially serialized with basic Python code blocks, such as for loops and list comprehensions. Specifically in this paper, we implemented operator resolution for element-wise operators, such as ReLU and element-wise addition. The computation of each entry of the input tensor is independent, so we can unroll the operator along one axis. For example, suppose the tensor $\mathbf{a}$ and $\mathbf{b}$ both have a shape of $d_1 \times d_2 \times d_3$ in the assignment statement $\mathbf{b} = ReLU(\mathbf{a})$. We first randomly choose a dimension as the axis of the resolution, and unroll the operator in an equivalent form using loops or list comprehensions. Taking the second dimension as an example, the expression can be expanded as:

\begin{Verbatim}[frame=single,commandchars=\\\{\},codes={\catcode`\$=3\catcode`\^=7\catcode`\_=8\relax}]
\PY{n}{b} \PY{o}{=} \PY{n}{torch}\PY{o}{.}\PY{n}{empty\PYZus{}like}\PY{p}{(}\PY{n}{a}\PY{p}{)}
\PY{k}{for} \PY{n}{i} \PY{o+ow}{in} \PY{n+nb}{range}\PY{p}{(}\PY{n}{a}\PY{o}{.}\PY{n}{shape}\PY{p}{[}\PY{l+m+mi}{1}\PY{p}{]}\PY{p}{)}\PY{p}{:}
    \PY{n}{b}\PY{p}{[}\PY{p}{:}\PY{p}{,} \PY{n}{i}\PY{p}{,} \PY{p}{:}\PY{p}{]} \PY{o}{=} \PY{n}{torch}\PY{o}{.}\PY{n}{relu}\PY{p}{(}\PY{n}{a}\PY{p}{[}\PY{p}{:}\PY{p}{,} \PY{n}{i}\PY{p}{,} \PY{p}{:}\PY{p}{]}\PY{p}{)}
\end{Verbatim}

\item \textbf{Mutate-then-recover}. The straight-line code converted from computational graphs is in the static single-assignment form where each variable will be assigned only once and remain constant, while real-world code often involves the mutation of values of variables. 
To introduce tensor value mutation while preserving the numerical equivalence, we adopt the mutate-then-recover strategy, as shown in Figure~\ref{fig:mtr}. 
We randomly choose a tensor and an entry of it, back up its value in a temporary variable, and modify its value with a random value. 
Then we scan the following code to find the next statement which depends on the chosen tensor, and we insert a statement to recover its value. Therefore, the final output of the mutated program would not be changed.

\item \textbf{Functionalization}. Python supports nested functions and closures. To test this feature, we randomly choose a sequence of operations in the program and wrap them into a function, as shown in \ref{fig:subfunc}. All variables defined or mutated in the function will be returned as reflected in the upper-level scope.
Since Python supports hoisting (using a variable defined after the definition of a function), we can move the constructed function anywhere before it is called.
\end{enumerate}

\subsection{Equivalence Modulo Inputs (EMI)}

\begin{figure}[t]
    \centering
    \includegraphics[width=0.99\textwidth]{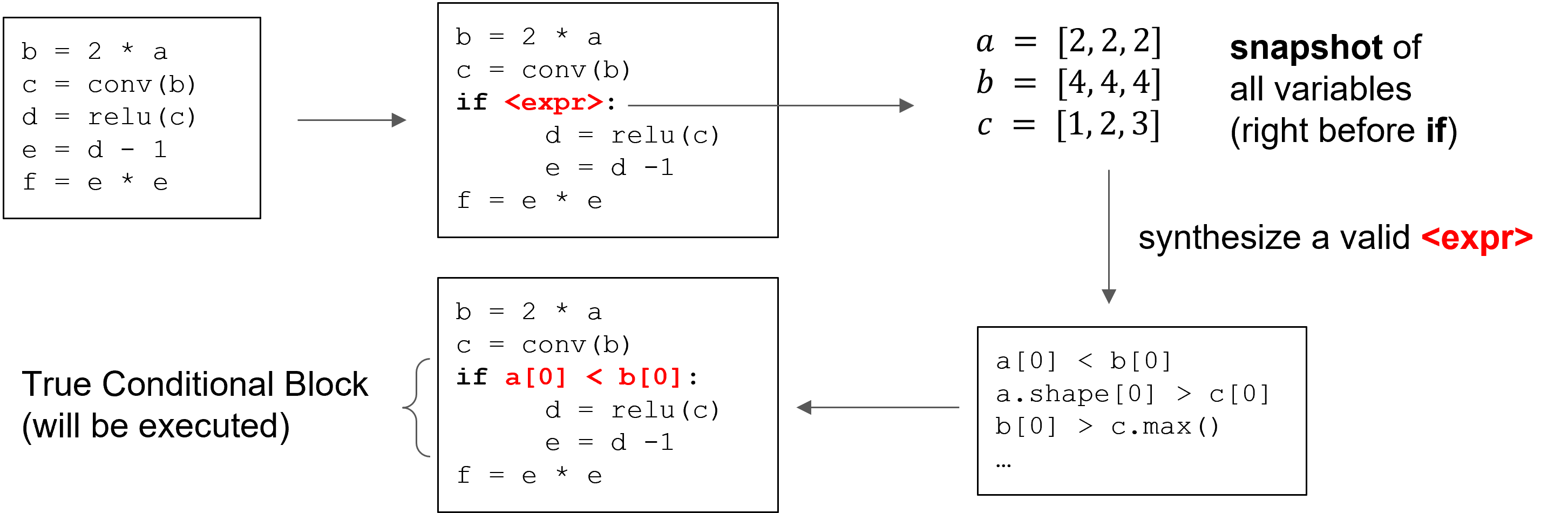}
    \caption{By leveraging program profiling  (runtime information that depends on $x_0$) and condition synthesis, Equivalence Modulo Inputs (EMI) ensures that mutated programs execute in the same order as the original programs. }
    \label{fig:emi}
\end{figure}

The second mutation method guarantees conditional program equivalence that depends directly on the specific given program input $x_0$, which is called \emph{Equivalance Modulo Inputs} (EMI). We want the mutated program gives the same result as the seed program on the given inputs, that is 
$$
\sema{\PP}(x_0) = \sema{\PP'}(x_0)
$$

Leveraging the fixed inputs enlarge the range of possible mutations. A class of EMI program mutations is \emph{Always True Conditional Block} (TCB). As shown in Figure ~\ref{fig:emi}, we can randomly select a subset of consecutive statements and wrap them up with an \texttt{if} statement. 
As long as the condition expression (highlighted as red in Figure~\ref{fig:emi}) is true, the wrapped statements will be executed as the original program, which is supposed to lead to identical outputs. Synthesizing a true condition expression takes two steps, 1) program profiling, where we collect the runtime information which depends on $x_0$, and 2) condition synthesis.

\paragraph{Program Profiling} We execute the original program step by step until the position to insert the \texttt{if} statement while maintaining a symbol table to book-keep the value of each tensor variable. Note that it is possible that the program is not in SSA form after other mutations, that is the value of a variable might be changed. In order to support the composition of mutations, we need to stop the profiling exactly before the \texttt{if} statement. 

\paragraph{Condition Synthesis} Following the grammar defined by Figure~\ref{fig:bnf}, we synthesize an arithmetic comparison expression based on the execution profile we collect in the previous step. The comparison expression takes two scalars as operands, which are either an element of a tensor, the length of a certain dimension of a tensor,  the number of dimensions of a tensor, or the maximum or minimum of a tensor. Since the input $x_0$ is fixed, all of these values are determined, and so is the order between two operands. Therefore the operator can also be determined.

\setlength{\grammarindent}{7em}

\begin{figure}\centering
\begin{subfigure}{0.7\linewidth}
\begin{grammar}
<cond> ::= <scalar> <op> <scalar>

<scalar> ::= <v> `[' <pos> `]' 
\alt <v> `.shape[' <pos> `]' | `len(' <v> `.shape)'
\alt <v> `.max()' | <v> `.min()' | <c>

<v> ::= variables

<c> ::= $\mathbb{Z}$

<pos> ::= <c> | <pos> `,' <c>

<op> ::= `>' | `<' | `>=' | `<='
\end{grammar}
\end{subfigure}
\caption{BNF Grammar of Synthesized Expressions}
\label{fig:bnf}
\end{figure}

\subsection{Mutations Beyond Equivalence}

It's noteworthy that the equivalence-based program mutations mentioned are only one possible approach for generating valid test cases. Their primary purpose is to ensure the meaningfulness of test cases, aligning them with real-world programs developed by machine learning engineers.

However,  it is important to clarify that equivalence is not a strict requirement for machine learning compiler fuzzing. Since we assume the interpreter always produces the correct results, the correctness of the compiled program can be verified by comparing it with the ground truth generated by the interpreter, as long as the test case is valid. Valid test cases are not necessarily derived from a seed program using equivalent mutations, and there can be other mutations that are not based on equivalence, which have the potential to enhance testing coverage. Nevertheless, how to design such mutations is beyond the scope of this work and remains a subject for future research.
\section{Evaluation}

\subsection{Testing Settings}
We implemented \sys in approximately 1000 lines of Python code on the top of NNSmith, including 4 types of program mutations mentioned in Section~\ref{sec: appr}.

We ran \sys on a workstation equipped with an NVIDIA RTX 2070 GPU. The operating system is Ubuntu 20.04. We build PyTorch from source in the latest main branch of PyTorch project on GitHub. Both CPU and GPU backends of the PyTorch compiler are tested.

We use NNSmith~\cite{nnsmith} as the seed program generator. Note that \sys is orthogonal to the seed program generator. Any fuzzing tools for DL systems based on computational graphs can be used as the seed program generator for \sys, and any improvements (e.g. supporting more operators) could benefit \sys. 

We set the number of operators in one computational graph as 20. Computational graphs that are too large will slow down the testing process and tend to trigger commonly seen bugs more frequently, impeding the discovery of new bugs in corner cases. Graphs with too few operators cannot cover a wide range of combinations of operators.

\begin{table}[h]
    \centering
    \begin{tabular}{ | c | c | c | c | }
    \toprule
        GitHub Issue ID & Crashed Component & Category & Fixed \\
    \midrule
        96432 & CPU Backend & Inconsistent Results & Yes \\
        96446 & CPU Backend & Compiler Crash & Yes \\
        96484 & CPU Backend & Compiler Crash & Yes \\
        96487 & CPU \& GPU Backend & Compiler Crash & No \\
        96604 & GPU Backend & Inconsistent Results & Yes \\
        96609 & GPU Backend & Compiler Crash & No \\
        96625 & CPU \& GPU Backend & Compiler Crash & Yes \\
        96728 & CPU Backend & Inconsistent Results & Yes \\
        97081 & Dynamo & Compiler Crash & No \\
        97082 & Dynamo & Compiler Crash & No \\
        97083 & CPU \& GPU Backend & Inconsistent Results & No \\
        97115 & Dynamo & Compiler Crash & No \\
        97117 & GPU Backend & Inconsistent Results & Yes \\
        97124 & CPU Backend & Compiler Crash & Yes \\
        97127 & CPU Backend & Compiler Crash & Yes \\
        97130 & Dynamo & Inconsistent Results & No \\
        97807 & GPU Backend & Compiler Crash & No \\
        
    \midrule
        1328 & Triton & Compiler Hang & Yes \\
        1337 & Triton & Inconsistent Results  & Yes \\
        1342 & Triton & Inconsistent Results & Yes \\
    \bottomrule
    \end{tabular}
    
    \caption{We find 17 bugs in PyTorch’s main branch and 3 bugs in PyTorch’s underlying GPU backend compiler. }
    \label{tab:bugs}
\end{table}

\subsection{Quantitative Results}

\sloppy
All bugs have been minimized and de-duplicated. We only list bugs that have distinct root causes in this section.
 The related GitHub issues can be found in \url{https://github.com/pytorch/pytorch/issues/created_by/soodoshll} and \url{https://github.com/openai/triton/issues/created_by/soodoshll}.

\paragraph{Bug Counts} Within one month, we found 17 bugs in PyTorch's main branch and three PyTorch's underlying GPU backend compiler Triton. All bugs have been minimized and reported to the community, and we list them in Table \ref{tab:bugs}. These bugs have been confirmed by the community and 12 of them have already been fixed. Five test cases have been integrated as unit tests of the main branch.

\paragraph{Bug Types} We discovered different types of bugs in multiple layers of the software stack. According to which part of the software the bugs happen in, they can be classified into:
\begin{itemize}
    \item TorchDynamo: The graph capturer fails to capture computational graphs. We find 4 such bugs and all of them are discovered by composite program mutations involving dynamic language features.
    \item CPU backend: The CPU backend of TorchInductor fails to generate the correct code. We found 9 such bugs (3 of which also cause errors in the GPU backend), which are the most among all categories. Type casting and operator fusion are two fragile parts.
    \item GPU backend: The GPU backend of TorchInductor fails to generate the correct code. We found 7 such bugs. %(3 of which also cause errors in the CPU backend).
    \item Triton: Bugs related to the underlying Triton compiler, which is maintained by another community. We found three bugs in Triton, and a fundamental one of them is that Triton evaluates the expression \texttt{True < False} as true. 
\end{itemize}
According to the error type the bugs, they can be categorized into:
\begin{itemize}
    \item Compiler Crash: The compiler crashes without producing runnable code.
    \item Compiler Hang: We found one bug in the Triton compiler where the compilation never terminates.
    \item Inconsistent Results: The results given by the original and compiled programs are different.
\end{itemize}

\subsection{Sample bugs}
We select some typical bugs to demonstrate the efficacy of different program mutations.

\begin{figure}
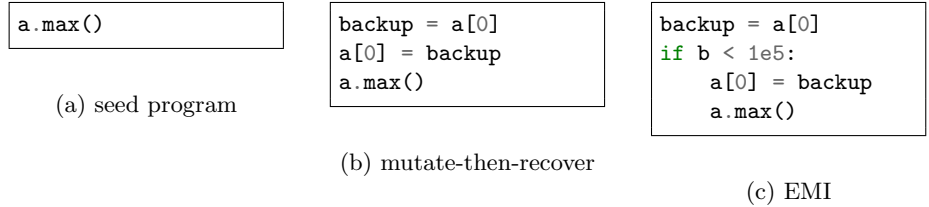
\centering

\begin{subfigure}[t]{0.3\textwidth}
\input{code/e1-a.py}
\caption{seed program}
\end{subfigure}
\hfill
\begin{subfigure}[t]{0.3\textwidth}
\input{code/e1-b.py}
\caption{mutate-then-recover}
\label{fig:e1-b}
\end{subfigure}
\hfill
\begin{subfigure}[t]{0.3\textwidth}
\input{code/e1-c.py}
\caption{EMI}
\end{subfigure}
\caption{Mutate-then-recover + EMI. The three snippets show how to derive new test cases step by step via mutations. This test case will cause TorchDynamo to crash due to an error in AOT Autograd (ahead-of-time auto-differentiation). The root cause is still under investigation.}
\end{figure}

\paragraph{Example 1: Mutate-then-recover + EMI.} Here {\tt b} is a scalar parameter of the upper-level function. In the first step, as shown in Figure~\ref{fig:e1-b}, we insert a mutate-then-recover snippet before the {\tt max} statement (the mutation statement is omitted since it will not affect the bug).  In the second step, we insert a TCB statement between the two statements we inserted in the last step. This example causes TorchDynamo to crash, and the root cause is still under investigation by the developers. This example demonstrates how the interaction between two different types of program mutation exposes undiscovered bugs.

\paragraph{Example 2: Mutate-then-recover + EMI + functionalize. } Figure~\ref{fig:e2} shows an example where three different types of mutation collaboratively triggered an error. It first creates three variables sharing the same underlying storage, i.e. they point to the same address. Therefore, any mutation of any one of the three tensors will also manifest in the other two. More specifically, this example modifies one element of the tensor {\tt b} from $1$ to $2$. So the value of tensor {\tt b} and {\tt c} is supposed to change accordingly, which is the behavior of the original code before compilation. The compiled code, however, incorrectly returns the {\tt c} with the unchanged value.

\begin{figure}[t]\centering
\input{code/e2.py}
\caption{Mutate-then-recover + EMI + Functionalize. The variables {\tt a}, {\tt b}, and {\tt c} are pointed to the same underlying memory address, so any mutation to one tensor should manifest on the other two. However, the compiled code does not behave as expected. The root cause of this bug is that TorchInductor does not handle in-place copy (\texttt{copy\_}) well.}
\label{fig:e2}
\end{figure}

% \qd{Following part needs to be rewritten}
\paragraph{Example 3: A bug caused by hoisting } Figure~\ref{fig:e3} shows an \textit{Pythonic} example, which one of the core developers of PyTorch compiler refers to as an `\emph{interesting puzzle}' and `\emph{a lot of fun for a PL nerd}'. This example is generated by three transformations collaboratively and has been minimized manually. It involves dynamic language features including nested functions, closure and variable hoisting.

\begin{figure}[t]
\centering
\input{code/e3.py}
\caption{A bug caused by hoisting. The function \texttt{subfunc} uses a variable \texttt{backup} in the upper-level scope, which appears after the definition of \texttt{subfunc}. PyTorch compiler crashes on this code snippet.}
\label{fig:e3}
\end{figure}
% \qd{TODO: Mutation Cost and Throughput}
% \section{Threats to Validity}
% \label{sec: threats}
\section{Related Work}
\label{sec: related}

\subsection{Compiler Testing}
Testing has been the dominating technique to validate the robustness of compilers. As an alternative to manually write test cases, automatic testing has been introduced to improve the testing coverage more efficiently. CSmith~\cite{csmith} gains great success in automatically finding bugs in C compilers including GCC and LLVM. It synthesizes programs from scratch and is used to generate seed programs by other mutation-based fuzzers. Hermes~\cite{sun2016finding} introduces several novel EMI mutations for live code regions and TCB is one of them. We borrow the idea from Hermes and extend it to Python and tensor operations.

% \paragraph{CSmith} CSmith~\cite{csmith} gains great success in automatically finding bugs in C compilers including GCC and LLVM. It synthesizes programs from scratch and is used to generate seed programs by other mutation-based fuzzers.

% \paragraph{EMI} Hermes~\cite{sun2016finding} proposes several novel EMI mutations for live code regions and TCB is one of them. We borrow the idea from Hermes and extend it to Python and tensor operations.

\subsection{Fuzzing DL Systems}

Fuzzing has been applied to deep learning systems to automatically detect bugs and improve their robustness. CRADLE~\cite{cradle} leverages existing DL models to run differential testing on deep learning systems. AUDEE~\cite{audee} and LEMON~\cite{lemon} extend CRADLE with mutation-based search strategies to improve the testing coverage. GraphFuzzer~\cite{graphfuzzer} and Muffin~\cite{graphfuzzer} enlarge the search space using the \texttt{reshape} operators. NNSmith~\cite{nnsmith}, which \sys is built upon, adopts SMT-solving to generate valid computational graphs and gradient-based search to find valid inputs, which further enlarge the search space. These works all focus on the computational graph level, while our work can generate test cases that cannot be represented by computational graphs. Furthermore, any graph-level fuzzers can be used as the seed program generator of \sys.

Besides graph-level fuzzers, there are also fuzzers like TVMFuzz \cite{tvmfuzz} and Tzer \cite{liu2022coverage} designed for lower-level tensor program compilers such as TVM. These tensor compilers correspond to the backends of PyTorch compilers and therefore the fuzzers targeting on them are not aware of the high-level dynamic language features. Other techniques like Predoo \cite{zhang2021predoo}, FreeFuzz \cite{wei2022free}, and DeepREL \cite{deng2022fuzzing} are helpful for testing DL operators, they are insufficient for identifying bugs in graph-level optimizations.

\subsection{Verified Compilers}
Verified compilers ensure their correctness by formal proofs. Liu \etal~\cite{liu2022verified} implemented a verified Coq framework for optimizing tensor programs written in a purely functional language with a set of verified program rewrites. For traditional compilers, CompCert~\cite{leroy2006formal} is a verified compiler for the C language. While formal verification can guarantee the correctness of compilers, it requires expertise in theorem proving and heavy manual intervention. What makes it worse is that programming languages and compilers for DL are still rapidly evolving, demanding extra efforts to update the corresponding proofs.

\subsection{Translation Validation}
Translation validation is another method to validate compiler optimizations, which verifies the compiled code is equivalent to the  source code. Unlike verified compilers, translation validation checks the equivalence between the target and source code of a specific input program. Bang~\etal~\cite{bang2022smt} uses SMT solver to verify the behavior of code compiled by the tensor program compiler MLIR is identical to the source code. It leverages to lower the complexity of high dimensional tensor data. There are also studies of applying translation validation on traditional languages like assembly~\cite{samet1975automatically}, C~\cite{siegel1998translation,necula2000translation}, Java~\cite{tate2009equality}, and LLVM~\cite{lopes2021alive2,tristan2011evaluating,kasampalis2021language,stepp2011equality}.
\section{Conclusions}
\label{sec: conclusion}

The introduction of PyTorch 2.0 signifies a remarkable milestone in the evolution of deep learning infrastructure by aiming to incorporate compiler techniques in a more dynamic manner. Nevertheless, the dynamic nature of Python poses challenges to the compiler's completeness and robustness.

While recent research has introduced fuzzing as a method to test deep learning compilers, there is still a lack of investigation concerning the testing of dynamic features. To bridge this gap, our proposed approach suggests multiple code transformations to generate test cases involving dynamic features. These transformations ensure the preservation of the program's semantics, thereby indicating the presence of a bug if any discrepancies arise between the transformed and original programs. Through this approach, we have identified a total of twenty bugs in the PyTorch compiler and its underlying tensor compiler Triton.

% Moreover, we identify two potential directions for future work. First, we propose enriching the set of program transformations by introducing more in-place operators, such as \texttt{add\_}, which have been found to be fragile components of deep learning frameworks. Second, we suggest making debugging more user-friendly by providing more useful hints. Currently, the process of minimizing test cases requires significant engineering efforts and manual intervention. Exploring ways to automate this process would be an intriguing avenue for further investigation.

% We also identify two directions of future work: 1) Enrich the set of program transformations. More in-place operators like \texttt{add\_} can be introduced since it has been a fragile part of DL frameworks; 2) Make debugging more friendly to developers by giving more useful hints. Now we still need to minimize the test case manually, which requires substantial engineering efforts. It would be interesting to investigate how to automate this process.

% We are continuing to run our fuzzer and trying to detect more bugs in PyTorch. Future works include:
% \begin{enumerate}
%     \item Enrich the set of program transformations. More in-place operators like \texttt{add\_} can be introduced since it has been a fragile part of DL frameworks.
%     \item Make debugging more friendly to developers by giving more useful hints. Now we still need to minimize the test case manually, which requires substantial engineering efforts. It would be interesting to investigate how to automate this process.
% \end{enumerate}

\section*{Acknowledgment}
We thank the anonymous reviewers for their insightful comments. This work was supported, in part, by Individual Discovery Grants from the Natural Sciences and Engineering Research Council of Canada and the Canada CIFAR AI Chair Program.
\bibliographystyle{splncs04}
\bibliography{ref}

\end{document}